\documentclass[usenatbib,twocolumn]{mn2e}
\usepackage{times}
\usepackage{graphicx}

\def\prd{Phys.Rev.D}

\pubyear{2008}

\author[Bagla and Prasad]
{J.~S.~Bagla and Jayanti Prasad \\
  Harish-Chandra Research Institute,  Chhatnag Road, Jhusi, Allahabad 211019,
  India. \\
  E-mail: (jasjeet, jayanti)@hri.res.in} 

\title[Gravitational collapse and the role of substructure]
{Gravitational collapse in an expanding background and the role of
  substructure II: Excess power at small scales and its effect 
  of collapse of structures at larger scales}  

\label{firstpage}

\def\LaTeX{L\kern-.36em\raise.3ex\hbox{a}\kern-.15em
    T\kern-.1667em\lower.7ex\hbox{E}\kern-.125emX}

\pagerange{\pageref{firstpage}--\pageref{lastpage}} 

\begin{document}

\maketitle

%%%%%%%%%%%%%%%%%%%%%%%%%%%%%%%%%%%%%%%%%%%%%%%%%%%%%%%%%%%%%%%%%%%%%%%

\begin{abstract}
We study the interplay of clumping at small scales with the collapse
and relaxation of perturbations at larger scales using N-Body simulations.   
We quantify the effect of collapsed haloes on perturbations at larger scales
using two point correlation function, moments of counts in cells and 
mass function.  
The purpose of the study is twofold and the primary aim is to quantify the
role played by collapsed low mass haloes in the evolution of perturbations at
large scales, this is in view of the strong effect seen when the large scale
perturbation is highly symmetric. 
Another reason for this study is to ask whether features or a cutoff in the
initial power spectrum can be detected using measures of clustering at scales
that are already non-linear.
The final aim is to understand the effect of ignoring perturbations at scales
smaller than the resolution of N-Body simulations. 
We find that these effects are ignorable if the scale of non-linearity is
larger than the average inter-particle separation in simulations. 
Features in in the initial power spectrum can be detected easily if the scale
of these features is in the linear regime, detecting such features becomes
difficult as the relevant scales become non-linear. 
We find no effect of features in initial power spectra at small scales on the
evolved power spectra at large scales.
We may conclude that in general, the effect on evolution of perturbations at
large scales of clumping on small scales is very small and may be ignored in
most situations.
\end{abstract}

%%%%%%%%%%%%%%%%%%%%%%%%%%%%%%%%%%%%%%%%%%%%%%%%%%%%%%%%%%%%%%%%%%%%%%%


%%%%%%%%%%%%%%%%%%%%%%%%%%%%%%%%%%%%%%%%%%%%%%%%%%%%%%%%%%%%%%%%%%%%%%%
%%%%%%%%%%%%%%%%%%%%%%%%%%%%%%%%%%%%%%%%%%%%%%%%%%%%%%%%%%%%%%%%%%%%%%%

\section{Introduction}

Large scale structures like galaxies and clusters of galaxies are
believed to have formed by gravitational amplification of small
perturbations
\citep{1980lssu.book.....P,1989RvMP...61..185S,1999coph.book.....P,2002tagc.book.....P,2002PhR...367....1B}.   
Much of the matter in galaxies and clusters of galaxies is the so
called dark matter \citep{1987ARA&A..25..425T,2003ApJS..148..175S} that is
believed to be essentially non-interacting and non-relativistic. 
The dark matter responds mainly to gravitational forces and by virtue of
larger density than the ordinary or Baryonic matter, the assembly of matter
into haloes and the large scale structure is driven by gravitational
instability of initial perturbations. 
Galaxies are believed to form when gas in highly over-dense haloes
cools and collapses to form stars in large numbers
\citep{1953ApJ...118..513H,1977MNRAS.179..541R,1977ApJ...211..638S,1977ApJ...215..483B}.   
Evolution of density perturbations due to gravitational interaction in a
cosmological setting is, therefore, the key process for the study of large
scale structure and its evolution and a very important one in formation and
evolution of galaxies. 
The basic equations for this are well known \citep{1974A&A....32..391P} and are
easy to solve when the amplitude of perturbations is small. 
At this stage, perturbations at each scale evolve independently on
perturbations at other scales and mode coupling is sub-dominant.  
Once the amplitude of perturbations at relevant scales becomes large the
coupling with perturbations at other scales becomes important and cannot be
ignored.
The equation for evolution of density perturbations cannot be solved for
generic perturbations in this regime, generally called the non-linear regime. 
One can use dynamical approximations for studying mildly non-linear
perturbations
\citep{1970A&A.....5...84Z,1989MNRAS.236..385G,1992MNRAS.259..437M,1993ApJ...418..570B,1994MNRAS.266..227B,1995PhR...262....1S,1996ApJ...471....1H,2002PhR...367....1B}.
Statistical approximations and scaling relations can be used if a limited
amount of information is sufficient
\citep{1977ApJS...34..425D,1991ApJ...374L...1H,1995MNRAS.276L..25J,2000ApJ...531...17Ka,1998ApJ...508L...5M,1994MNRAS.271..976N,1996ApJ...466..604P,1996MNRAS.278L..29P,1996MNRAS.280L..19P,2003MNRAS.341.1311S}. 
In general, however, we require cosmological N-Body simulations
\citep{1998ARA&A..36..599B,2004astro.ph.11043B} to follow the detailed
evolution of the system. 

In N-Body simulations, we simulate a representative region of the
universe.
This region is a large but finite volume. 
Effect of perturbations at scales smaller than the mass resolution of
the simulation, and of perturbations at scales larger than the box is
ignored.
Indeed, even perturbations at scales comparable to the box are under
sampled.  
It has been known for a long time that perturbations at scales much larger
than the simulation volume can affect the results of N-Body simulations
\citep{1994ApJ...436..467G,1994ApJ...436..491G,1996ApJ...472...14T,1997MNRAS.286...38C,2005MNRAS.358.1076B,2005astro.ph..3106S,2006MNRAS.370..993B,2006MNRAS.370..691P,2008arXiv0802.1808T,2008arXiv0804.1197B}. 
It is possible to quantify these effects and even estimate whether a given
simulation volume is large enough to be representative or not
\citep{2005MNRAS.358.1076B,2006MNRAS.370..993B}. 
It has been shown that for gravitational dynamics in an expanding
universe, perturbations at small scales do not influence collapse of
large scale perturbations in a significant manner
\citep{1974A&A....32..391P,1985ApJ...297..350P,1991MNRAS.253..295L,1997MNRAS.286.1023B,1998ApJ...497..499C}
as far as the correlation function or power spectrum at large scales are
concerned. 
This has led to a belief that ignoring perturbations at scales much
smaller than the scales of interest does not affect results of N-Body
simulations.  
Recently we have shown that if large scale collapse is highly symmetric then
presence  of perturbations at much smaller scales affect evolution of density
perturbations at large scales \citep{2005MNRAS.360..194B}.
Here we propose to study the effect of small scales on collapse of
perturbations at large scales in a generic situation.  

Substructure can play an important role in the relaxation process.
It can induce mixing in phase space
\citep{1967MNRAS.136..101L,2001MNRAS.328..311W}, or change halo profiles by
introducing transverse motions
\citep{1990ApJ...365...27P,2000ApJ...538..517S}, and, gravitational
interactions between small clumps can bring in an effective collisionality
even for a collisionless fluid \citep{2004PhRvL..93b1301M,2003astro.ph.11049M}.
Thus it is important to understand the role played by substructure in
gravitational collapse and relaxation in the context of an expanding
background. 

Whether the evolution of density perturbations is affected by collapsed
structure or not depends on the significance of mode coupling between these
scales. 
We summarize the known results about mode coupling here.
\begin{itemize}
\item
Large scales influence small scales in a significant manner.  
If the initial conditions are modified by filtering out perturbations at small
scales then mode coupling generates small scale power.  
If the scale of filtration is smaller than the scale of non-linearity at the
final epoch then the non-linear power spectrum as well as the appearance of
large scale structure is similar to the original case
\citep{1985ApJ...297..350P,1991MNRAS.253..295L,1997MNRAS.286.1023B,1998ApJ...497..499C}.  
\item
Non-linear evolution {\it drives} every model towards a weak attractor ($P(k)
\simeq k^{-1}$) in the mildly non-linear regime ($1 \leq \bar\xi \leq 200$)
\citep{1992ApJ...399..397K,1997MNRAS.286.1023B}. 
\item
In absence of initial perturbations at large scales, mode coupling generates
power with ($P(k) \simeq k^4$) that grows very rapidly at early times
\citep{1997MNRAS.286.1023B}.  There are a number of explanations for this
feature, ranging from second order perturbation theory to momentum and mass
conserving motion of a group of particles.  The $k^4$ tail can also be derived
from the full non-linear equation for density
\citep{1974A&A....32..391P,1980lssu.book.....P,1965AAA.....3...241Z}.
\item
If we consider large scale perturbations to be highly symmetric, e.g. planar,
then small scale fluctuations play a very important role in the non-linear
evolution of perturbations at large scales \citep{2005MNRAS.360..194B}. 
\end{itemize}

While the effect of large scales on small scales is known to be significant,
particularly if the larger scales are comparable to the scale of
non-linearity, the effect of small scales on larger scales is known to be
small in most situations.  
Even though this effect has not been studied in detail, many tools have been
developed that exploit the presumed smallness of the influence of small scales
on large scales
\citep{1996ApJS..103....1B,2002MNRAS.331..587M,2002ApJ...564....8M}. 

Considerable work has been done in recent years on the effects of the {\sl
  pre-initial}\/ conditions used in N-Body simulations \citep{2007PhRvE..75e9905B,2007PhRvE..75b1113B,2007PhRvE..76a1116B,1997Prama..49..161B,2006PhRvE..74b1110G,2007PhRvD..76j3505J,2007PhRvD..75f3516J,2005PhRvL..95a1304J,2006PhRvD..73j3507M}.  
We use the term pre-initial conditions to refer to the distribution of
particles on which the initial density and velocity perturbations are
imprinted.  
The pre-initial conditions are expected to have no density perturbations or
symmetry, but it can be shown that at least one of these requirements must be
relaxed in practice. 
This can lead to growth of some modes in a manner different from that expected
in the cosmological perturbation theory. 
Our work allows us to estimate the effect such discrepant modes can have on
the non-linear evolution of clustering at these scales.
Our work also allows us to understand the effects that may arise if the
primordial power spectrum deviates strongly from a power law at small scales. 

The evolution of perturbations at small scales depends strongly on the mass
and force resolution. 
A high force resolutions can lead to better modelling of dense haloes, but
gives rise to two body collisions
\citep{1998ApJ...497...38S,2002MNRAS.333..378B,2004MNRAS.348..977D,2004MNRAS.350..939B,2006MNRAS.370.1247E,2008arXiv0804.0294R}.
A high force resolution without a corresponding mass resolution can also give
misleading results as we cannot probe {\it shapes}\/ of collapsed objects
\citep{1996ApJ...470L..41K}.
In addition, discreteness and stochasticity also limit our ability to measure
physical quantities in simulations, and these too need to be understood
properly \citep{2008arXiv0803.3120T,2008arXiv0804.0294R}. 
In all such cases, the errors in modelling is large at small scales. 
It is important to understand how such errors may spread to larger scales and
affect physical quantities.

%%%%%%%%%%%%%%%%%%%%%%%%%%%%%%%%%%%%%%%%%%%%%%%%%%%%%%%%%%%%%%%%%%%%%%%
\begin{figure*}
\begin{center}
\begin{tabular}{cc}
\includegraphics[height=3.4truein]{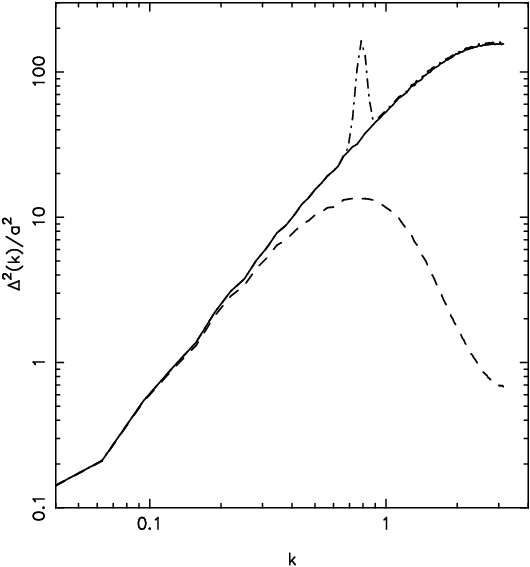}  &
\includegraphics[height=3.4truein]{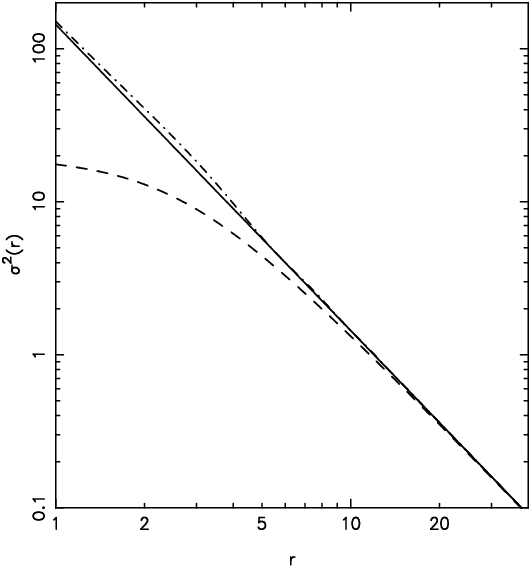} 
\end{tabular}
\caption{The left and right panels in this figure show the linearly
extrapolated power spectrum $\Delta^2(k)$ and  variance $\sigma^2(r)$
at the epoch $a=1$. In both the panels models I, II and III are 
represented by the solid, dashed and dot-dashed lines respectively.}
\end{center}
\label{init_pspc}
\end{figure*}
%%%%%%%%%%%%%%%%%%%%%%%%%%%%%%%%%%%%%%%%%%%%%%%%%%%%%%%%%%%%%%%%%%%%%%%

\section{Models}

In order to understand the effects of small scales perturbations on large
scales we have simulated some models numerically. In these models we either 
suppress or add extra power at small scales, with respect to our reference 
model (Model I). Some of the details of our cosmological simulations are as
follows:

\begin{itemize}
\item
The TreePM code \cite{2002JApA...23..185B,2003NewA....8..665B} for
cosmological N-Body simulations. 
\item
$200^3$ particles  in a volume of $200^3$ cubical cells for each simulation.
\item
A softening length of $0.5$ times the average inter-particle separation in
order to suppress two body collisions.
\item
$P(k) ~ = ~ A ~ k^{-1}$ as the reference model (Model I).  This was normalized
so that $\sigma^2(r=r_{nl},a=1)=1$ where $r_{nl}=12$ grid lengths. 
\item
Einstein-de-Sitter universe\footnote{Non-linear gravitational clustering is
  not likely to have a strong dependence on the choice of cosmology.  By
  restricting ourselves to the Einstein-de-Sitter universe, we have an
  additional check on simulations in form of self-similar evolution of
  clustering for the reference model.}.
\end{itemize}
We studied the following modifications of the reference spectrum.
\begin{itemize}
\item
Model II: Gaussian truncation of the reference power spectrum at small
scales. $P(k) ~ = ~ A ~ k^{-1} ~ \exp\left[ - {k^2}/{k_c^2} \right]$.  We
chose $k_c = k_{nyq}/4$, so that truncation is mainly at scales that are
smaller than the scale of non-linearity at late times.  A is chosen to be the
same as for Model~I. 
\item
Model III: A spike is added to the reference power spectrum.  $P(k) ~ = ~ A ~
k^{-1} + \alpha ~ A ~ k_c^{-1} \exp\left[ - \left(k-k_c\right)^2 / 2
  \sigma_k^2 \right]$.  We chose same $k_c$ as in Model~II, $\sigma_k = 2 \pi
/L_{box}$ is same as the fundamental wave number and we took $\alpha=4$.  A is
chosen to be the same as for Model~I.  
\end{itemize}
Clearly, these models have additional or truncated power at small scales as
compared to the reference model while large scales are the same in all the
models. 
The left and right panels in Figure~1 show the linearly extrapolated power 
spectrum $\Delta^2(k)$  which we start with in N-body simulations and the
theoretical mass variance $\sigma^2(r)$ respectively for the three  models
being considered at the last epoch  i.e., $a=1$.   
From both the panels of Figure~1 we  see that all the three 
models have identical power at the scales much larger than the scale at which 
we add or suppress the power i.e.,  $2\pi/k_c$. 

We choose to work with the Einstein-de Sitter universe as the background as
effects are mode coupling are more important in the non-linear regime and we
do not expect the cosmological parameters to influence the evolution of
perturbations at these scales. 
The specific choice of Einstein-de Sitter universe is useful as power law
initial conditions, e.g., the reference model, are expected to evolve in a
self similar manner and this provides a useful check for errors creeping in
due to the effects of a finite box-size or other numerical artifacts.

%%%%%%%%%%%%%%%%%%%%%%%%%%%%%%%%%%%%%%%%%%%%%%%%%%%%%%%%%%%%%%%%%%%%%%%
\begin{figure*}
\begin{center}
\begin{tabular}{cc}
\includegraphics[width=2.9truein]{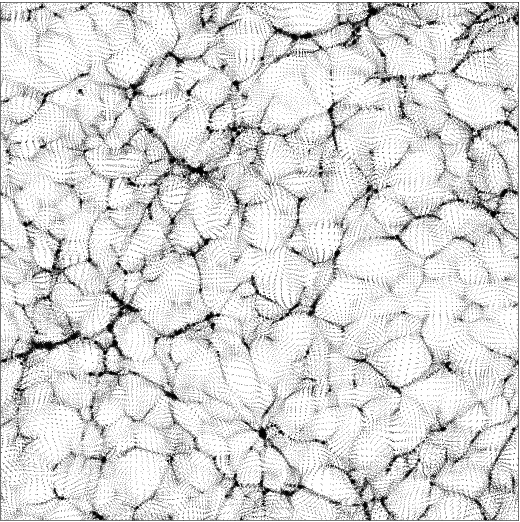} & 
\includegraphics[width=2.9truein]{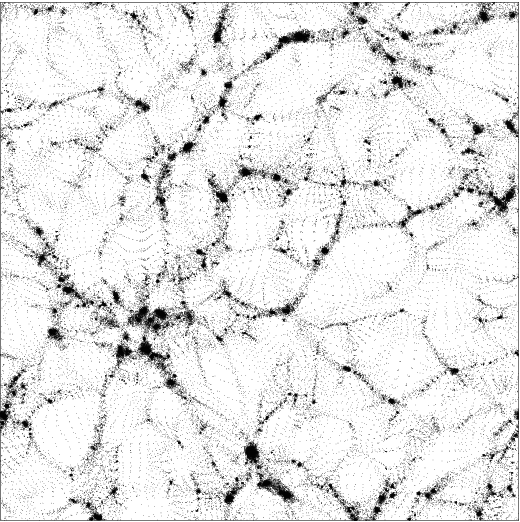} \\
\includegraphics[width=2.9truein]{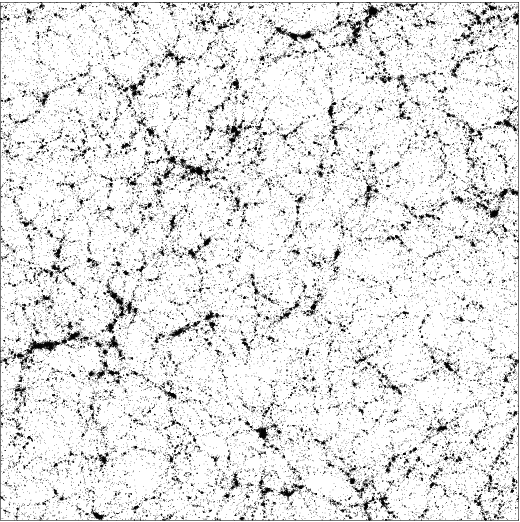} & 
\includegraphics[width=2.9truein]{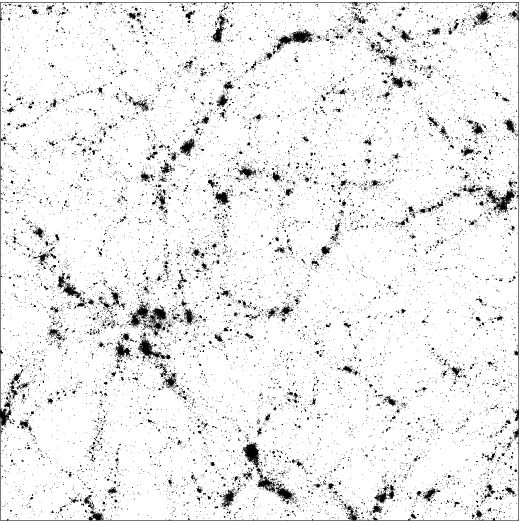} \\
\includegraphics[width=2.9truein]{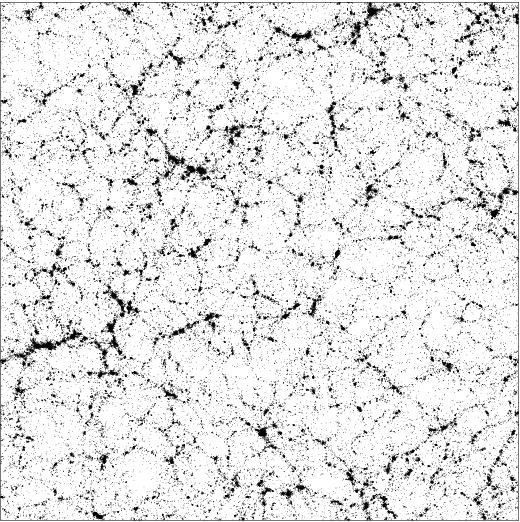} & 
\includegraphics[width=2.9truein]{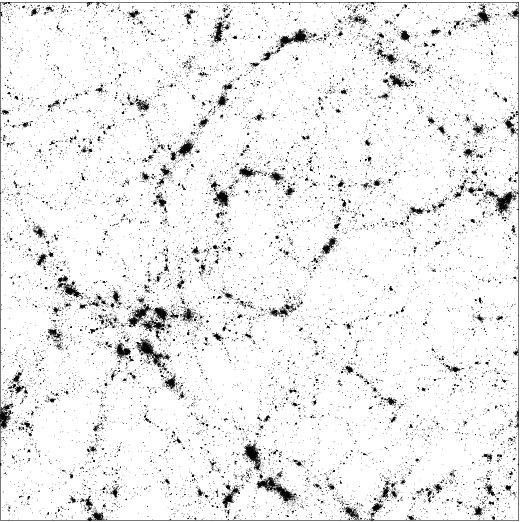} 
\end{tabular}
\end{center}
\caption{The first, second and third row in figure show the slices 
for the  model II, I and III respectively, at an early (first column)
and a later epoch (second column). The early epoch is identified with an 
epoch when the scale at which we add or truncate power i.e.,  $2\pi/k_c$,
 is linear in the  Model I and and the  later epoch is identified with an epoch
 when the scale $2\pi/k_c$ is  nonlinear in  Model II.}
\label{slices}
\end{figure*}
%%%%%%%%%%%%%%%%%%%%%%%%%%%%%%%%%%%%%%%%%%%%%%%%%%%%%%%%%%%%%%%%%%%%%%%

%%%%%%%%%%%%%%%%%%%%%%%%%%%%%%%%%%%%%%%%%%%%%%%%%%%%%%%%%%%%%%%%%%%%%%%
\begin{figure*}
\begin{center}
\begin{tabular}{cc}
\includegraphics[width=2.8truein]{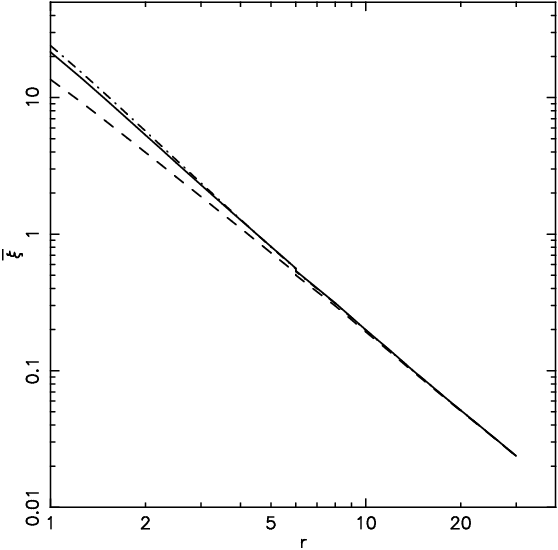} & 
\includegraphics[width=2.8truein]{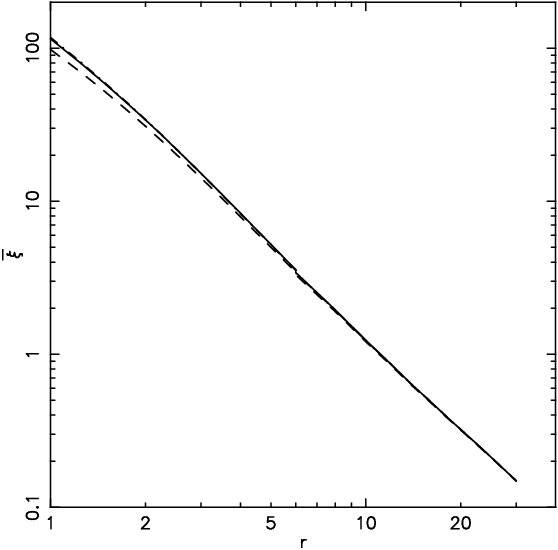} \\
\includegraphics[width=2.8truein]{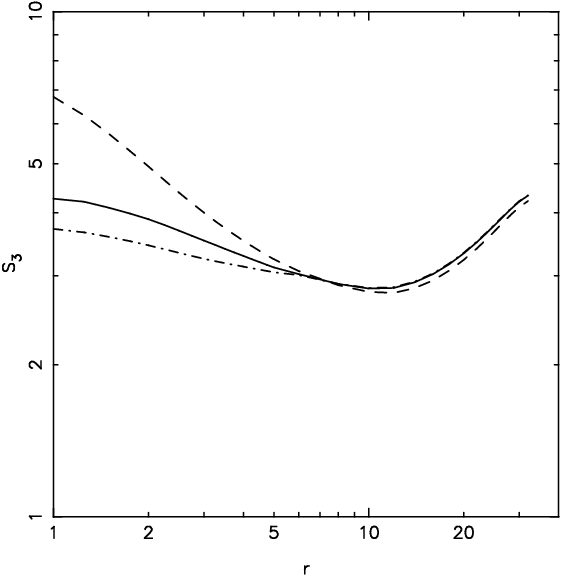} & 
\includegraphics[width=2.8truein]{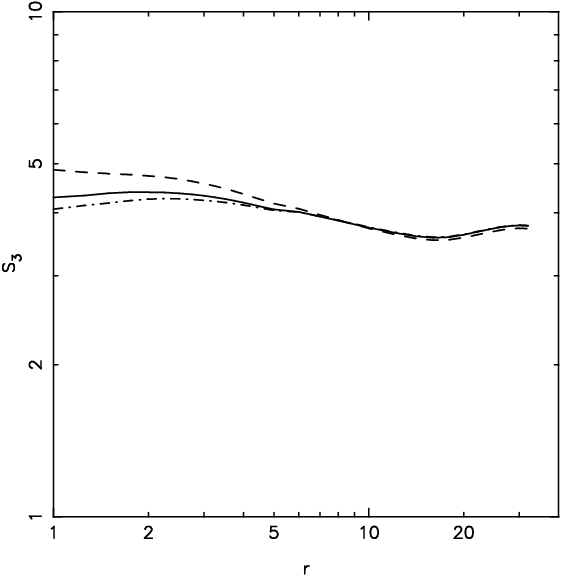} \\
\includegraphics[width=2.8truein]{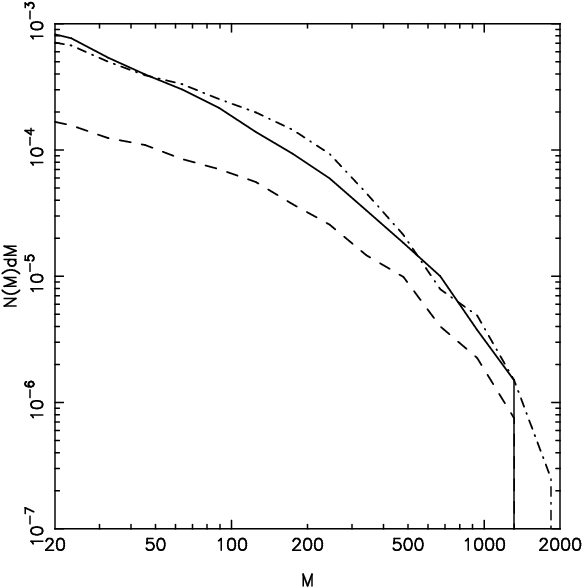} &
\includegraphics[width=2.7truein]{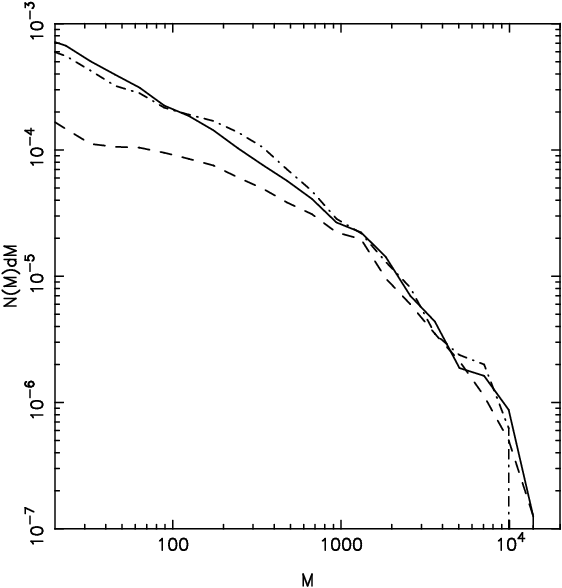} 
\end{tabular}
\end{center}
\caption{The first, second and third row in this figure show the average two
  point correlation function ${\bar \xi}$ , skewness $S_3$
  and comoving number density of haloes $N(M)dM$ respectively, 
  at an early (first column) and at a  later  epoch (second colum).
  Different models in all the panels are reprsented by the same line
  styles as in Figure~1.}
\label{clustering}
\end{figure*}
%%%%%%%%%%%%%%%%%%%%%%%%%%%%%%%%%%%%%%%%%%%%%%%%%%%%%%%%%%%%%%%%%%%%%%%

\section{Results}

Our goal is to understand the effects of variations in the initial power
spectrum at small scales on other scales.  
For this, we study the three models at two representative epochs: one where
the scale of modification is linear, and the second epoch when the
scale of non-linearity is larger than the scales where the power spectra
differ from each other. 
We refer to these epochs as an early epoch and a late epoch.  
The scale of non-linearity in the reference model at the early epoch is
$4.8$~grid lengths and the corresponding scale at the late epoch is $12$~grid
lengths.  
The wave number $k_c$ corresponds to $8$~grid lengths and becomes non-linear
at an intermediate epoch.

Figure~2 shows the distribution of particles in a thin slice from simulations
of the three models. 
The left column shows the distribution at the early epoch whereas the right
column shows the same slice at late times.  
The middle row shows the reference model (Model I), the top row is for model
with less power at small scales (Model II) and the bottom row is for the model
with excess power at small scales (Model III). 
The large scale distribution of particles is similar in all the three models
for both epochs, although there are significant differences at small scales. 
Differences are more prominent between model II and the other models, whereas
the differences between models I and III are less obvious.
Also, differences between the models diminish as we go from the early epoch to
the late epoch.

Figure~3 compares the models in a more quantitative manner. 
We have plotted the amplitude of clustering $\bar\xi(r)$ as a function of $r$
for the three models at an early epoch (top-left frame) and at a later epoch
(top-right frame).  
The differences between the amplitude of clustering are more pronounced at the
early epoch, though even here the differences are much smaller than those seen
in Figure~1 where the linearly extrapolated $\sigma^2(r)$ has been plotted. 
At late times, models I and III have an indistinguishable $\bar\xi(r)$ whereas
model II has a slightly smaller amplitude of clustering at small scales when
compared with these two models.
At very large $r$ compared to the scale of modification, all models have the
same $\bar\xi$ even at the early epoch.

Second row in figure~3 shows $S_3$ as a function of scale for the three
models.  
As before, the left panel is for the early epoch and the right panel is for
the late epoch. 
At large scales, larger than the scale of modification ($8$ grid lengths), the
three models agree well. 
There are significant differences at small scales, particularly at the early
epoch.  
Model II has the highest Skewness, whereas Model III has the smallest Skewness
at small scales. 
This ranking does not change with time, though the differences between models
decrease with further evolution of the system.

The bottom row in figure~3 shows the number density of haloes as a function of
mass.  
Mass here is shown in units of mass of each particle. 
Haloes were identified using the Friends-of-Friends algorithm (FOF) with a
linking length of $0.1$. 
We chose this linking length in order to avoid identifying smooth filaments in
model II as haloes. 
Haloes with a minimum of $20$ particles were considered for this plot.
We find that model III has the largest number of haloes around the
scale of modification, whereas model II has the least number of haloes at this
scale.   
Indeed, at the early epoch, model II has a much lower number of haloes at all
mass scales when compared with model I and model III. 
At late times, model II continues to have fewer small mass haloes though it
almost matches the other two masses at larger masses. 

We find that the two point correlation function does not retain any
information about differences in initial conditions after the scale where such
differences are present becomes sufficiently non-linear.  
This is in agreement with results of earlier studies
\citep{1985ApJ...297..350P,1991MNRAS.253..295L,1992ApJ...399..397K,1997MNRAS.286.1023B,1998ApJ...497..499C}. 

Skewness is a slightly better indicator than the two point correlation
function, in that it retains some information about the missing power at small
scales in model II even after the cutoff scale becomes non-linear.  
It does not retain much information about the excess power that is added at
small scales in model III.
One possible reason for this is that the cutoff affects the shape of the power
spectrum at $k \ll k_c$ whereas the effect of adding extra power is more
localized. 
We may conclude that Skewness is able to retain information about a cutoff in
the initial power spectrum if the cutoff scale is not strongly non-linear. 
This may not have implications for observational signatures of a cutoff as
observations of galaxy clustering are restricted to the redshift space and it
has been shown that redshift space distortions in the non-linear regime erase
differences between models \citep{2006astro.ph..4598B}.

The number density of haloes at scales comparable to and smaller than the
cutoff is smaller than that in the other models even after the cutoff scale
becomes non-linear. 
The mass function appears to be the most sensitive indicator of a cutoff in
the power spectrum in the mildly non-linear regime.

%%%%%%%%%%%%%%%%%%%%%%%%%%%%%%%%%%%%%%%%%%%%%%%%%%%%%%%%%%%%%%%%%%%%%%%

\section{Discussion}

We find that the memory of localized variations initial conditions is
erased in the quasi-linear regime.
This erasure is almost complete in measures of the second moment, e.g. the two
point correlation function shown here.  
We have checked that the same is true of the power spectrum and rms
fluctuations in mass. 
This loss of information has been pointed out in earlier work 
\citep{1991MNRAS.253..295L,1997MNRAS.286.1023B}. 
The Skewness appears to be a better indicator of a cutoff in the initial power
spectrum, at least in the quasi-linear regime.  
We find that the Skewness for model II is distinctly higher than that for
model I or model III, even when the scale of nonlinearity is much larger than
the cutoff scale. 
Number densities of haloes is a very faithful indicator of the cutoff, even at
late times. 
This is to be expected given that the number density of haloes can be
predicted fairly accurately using the Press-Schechter mass function
\citep{1974ApJ...187..425P} that relies only on the initial power spectrum. 

The question now arises as to how may we interpret these results.  
Here we would like to recall the key conclusion of paper~I in the present
series \citep{2005MNRAS.360..194B}. 
In paper~I, we studied the collapse of a plane wave with varying amount of
collapsed haloes at a much smaller scale (as compare to wavelength of the
plane wave).
We found that the thickness of pancake that forms due to collapse of the plane
wave is smaller if collapsed haloes are present.
The reason for smaller thickness is that gravitational interaction of
infalling clumps takes away some of the longitudinal momentum and leads to an
increase of the transverse momentum. 
Thinner pancakes imply a higher density, and clumps are able to grow very
rapidly in such an environment.
We were motivated to study collapse of a plane wave as it is known from the
Zel'dovich approximation \citep{1970A&A.....5...84Z} that locally, generic
collapse is planar leading to formation of pancakes. 

In case of generic initial conditions that we consider here, there is no fixed
large scale that is collapsing as we have perturbations at all scales. 
However, we have ensured that perturbations at large scales are the same in
all the three models. 
In this case the effect of power on large scales is to cause collapse around
peaks of density, or equivalently, empty the voids.
The latter picture is more attractive as it also tells us why collapse of
perturbations at a scale leads to enhancement of power at smaller scales
without any loss of power at the original scale: emptying under dense regions
simply puts more and more matter in thin walls around the void that forms.  
We can say that power is transferred from the scale of perturbation, that
essentially is the radius of the void that forms, to the scale of thickness of
pancakes surrounding the void. 
As a given scale becomes non-linear, we begin to see voids corresponding to
this scale. 
Matter that collapsed at an earlier stage gets pushed into the pancakes
surrounding the voids.
This information about the shape of the initial power spectrum at scales
smaller than the scale of non-linearity is mostly restricted to the
distribution of matter within pancakes. 
This, in our view, explains the erasure of memory of initial conditions for
the two point correlation function. 
The mass function and Skewness are more sensitive to the arrangement of matter
within pancakes and hence these remain different for the model with a cutoff. 
Once the scale of cutoff becomes strongly non-linear, most of the
perturbations at this scale are expected to be part of highly over dense
haloes. 
At this stage, we expect that all indicators of clustering will loose
information about the details of the initial power spectrum at this stage. 

In models I and III, there is significant amount of initial power at small
scales. 
This leads to a fragmented appearance of pancakes and clearly pancakes cannot
be thinner than the clumps.
In model II, there is no initial power at small scales. 
Power is generated at these scales by collapse of larger modes, power grows
very rapidly at small scales and the non-linear power spectrum in this case
catches up with the power spectrum for the other two models. 

Model III has significantly more power as compared with the reference model at
small scales. 
This leads to a more rapid growth of perturbations at these scales, as is seen
in the number density of collapsed haloes at the relevant scales at early
times. 
At late times, these haloes are assimilated into bigger haloes and we rapidly
loose any signatures of the excess power. 
We expect the excess power to lead to thinner pancakes, motivated by
conclusions of paper~I. 
However, the scale of pancakes is such that this feature is not apparent.

Another approach towards understanding the lack of effect of variations in
power spectrum at small scales on larger, non-linear scales is based on the
equation for evolution of density contrast \citep{1974A&A....32..391P}.  
It has been shown that the leading order effect of virialised haloes
on modes at much larger scales vanishes at the leading order.
In case of arbitrary motion of a group of particles, a $k^4$ tail is
generated in the power spectrum at $k \rightarrow 0$ if there is no initial
power at these scales. 
In general the influence of motions of particles at small scales to density
perturbations is limited due to the $k^2$ behaviour\footnote{A $k^2$
  dependence in density contrast translates into a $k^4$ dependence in the
  power spectrum.} of the mode coupling terms
in equations that describes the evolution of density contrast for a system of
particles \citep{1974A&A....32..391P}.  
The magnitude of the mode coupling at this order is proportional to the
departure from virial equilibrium for the system of particles.

It is possible to consider the equations and compute the leading order
contribution of interacting haloes. 
Clearly, this also must scale as $\mathcal{O}(k^2)$ for density contrast, but
it is instructive to see if we can quantify the level to which the internal
structure of clusters matters for coupling of density fluctuations. 
A detailed calculation and analysis of this is presented in a forthcoming
manuscript (Bagla, In Preparation).  
We summarise a few key points here.
\begin{itemize}
\item
It can be shown that the leading order contribution to mode coupling for
interacting haloes comes from the halo-halo interaction where the haloes may
be assumed to be point masses.
\item
There is a next to leading order contribution due to tidal interaction of
clusters. 
\item
The two contributions scale as $k^2$, making the contribution to power
spectrum as  $\mathcal{O}(k^4)$.
\item
The magnitude of the leading order term is proportional to the departure from
virial equilibrium for haloes treated as point masses.
\end{itemize}
The treatment can be generalised to an arbitrary number of haloes, and we can
also study the effects of coupling between a cluster and the large scale
density distribution. 
These conclusions explain the results of our numerical experiments, and give a
reason as to why gravitational clustering in an expanding universe appears to
be {\sl almost} renormalizable.

\section{Summary}

Results presented in the preceeding section show that for a hierarchical
model, there is little effect of features at small scales (high wave number)
in power spectrum on collapse of perturbations at larger scales. 
At the same time, we see that the effect of features can be seen in several
statistical indicators at the scales of features and also at smaller scales. 
The key conclusion that we can draw is that if we modify the power spectrum at
small scales, there is no discernable effect of these modifications at larger
scales. 
This has implications in several situations:
\begin{itemize}
\item
Cosmological N-Body simulations start with initial conditions that do not
sample the power spectrum at large wave numbers.  
In typical simulations of this type, a grid is used to generate initial
conditions and only modes up to the Nyquist wave number are sampled.  
Indeed, if the number of particles is smaller than the number of grid cells
used for generating initial conditions, the effective upper limit to wave
numbers is even more restricted \citep{1997Prama..49..161B}.
The missing part of the power spectrum does not have any impact on the
evolution of non-linear structures at scales larger than the cutoff scale. 
We expect the effects of missing modes at large wave numbers to be less and
less relevant as larger length scales (smaller wave numbers) become
non-linear. 
\item
It has been pointed out that the choice of pre-initial conditions, and the
epoch at which the initial conditions are set up can lead to spurious growth
of some modes
\citep{2007PhRvE..75e9905B,2007PhRvE..75b1113B,2007PhRvE..76a1116B,2006PhRvE..74b1110G,2007PhRvD..76j3505J,2007PhRvD..75f3516J,2005PhRvL..95a1304J,2006PhRvD..73j3507M}.   
Clearly, these effects must be suppressed as the modes with spurious growth
become non-linear. 
\item
Generation of perturbations in the early universe, and their evolution towards
the end of the inflationary phase can lead to a scale dependent evolution of
modes \citep{2004PhRvD..69f3505M,2007PhRvD..76d3530D}.
Our work clearly shows that such features will be impossible to detect if
these are at scales that are strongly non-linear and difficult to detect if
these are at scales that are mildly non-linear.  
If scales where such variations occur are already non-linear then these
variations do no affect collapse of larger scales. 
Of course, if the scales where such variations occur are linear then these can
be probed using galaxy clustering.
\end{itemize}

%%%%%%%%%%%%%%%%%%%%%%%%%%%%%%%%%%%%%%%%%%%%%%%%%%%%%%%%%%%%%%%%%%%%%%%

\section*{Acknowledgments}

Numerical experiments for this study were carried out at cluster computing
facility in the Harish-Chandra Research Institute
(http://cluster.mri.ernet.in).  
This research has made use of NASA's Astrophysics Data System.

%%%%%%%%%%%%%%%%%%%%%%%%%%%%%%%%%%%%%%%%%%%%%%%%%%%%%%%%%%%%%%%%%%%%%%%
%%%%%%%%%%%%%%%%%%%%%%%%%%%%%%%%%%%%%%%%%%%%%%%%%%%%%%%%%%%%%%%%%%%%%%%

\label{lastpage}

\end{document}